\documentclass[11pt]{article}
\usepackage{hyperref}
\pdfoutput=1

\begin{document}
\title{Influence of Liquid Viscosity on Droplet Impingement \\ on Superhydrophobic Surfaces}
\author{John T. Pearson \\ Daniel Maynes \\ Brent W. Webb
\\\vspace{6pt} Department of Mechanical Engineering
\\ Brigham Young University, Provo, UT 84602, USA}
\maketitle

\begin{abstract}
This fluid dynamics video describes droplet impingement experiments performed on superhydrophobic surfaces.  When droplets of pure water are impinged upon superhydrophobic surfaces, a region of thin coherent jets are observed for Weber numbers between 5 and 15.  Also, peripheral splashing is observed for Weber numbers above about 200.  When the viscosity of the droplet is increased by mixing glycerol with the water, the thin jets are not observed and peripheral splashing is delayed somewhat.  In the Weber number range where pure water droplets are observed to splash peripherally, the water/glycerol droplets are observed to have two-pronged jets.
\end{abstract}

% main text

\section{Video Description}

This video describes the influence of liquid viscosity on droplet impingement experiments performed on superhydrophobic surfaces.  The surfaces used in these experiments are comprised of alternating micro-ribs and cavities which have been coated with a thin Teflon coating.  When a droplet impinges upon a superhydrophobic surface, it spreads, retracts, and ejects a vertical jet.  The focus of the video is on the dynamics associated with the ejected jet.
\\The jet dynamics are Weber number dependent.  The Weber number is a measure of the importance of inertial forces compared to surface tension forces.  For Weber numbers between 5 and 15, the formation of thin coherent jets is observed when pure water is the fluid type.  These thin jets have a very high vertical velocity, sometimes exceeding 15 times the impact velocity, as shown in the video.  When Weber numbers exceed about 200, the pure water droplets are observed to splash peripherally.  This results in a much weaker jet.
\\When the viscosity of the fluid is increased, the jet dynamics are significantly altered.  For a mixture of 50 percent water and 50 percent glycerol, the viscosity is six times that of pure water, and the surface tension is about the same.  For this fluid type, the thin jets are entirely suppressed.  Also, the peripheral splashing is delayed somewhat.
\\One very interesting phenomenon observed with this fluid type is the formation of two-pronged jets for Weber numbers between 150 and 300.  The inward velocities during retraction are unequal, resulting in an ejecting jet that is non-circular.  This is an effect of the micro-ribs on the surface.  When the Teflon coating is removed the two-prongs are still observed, though only immediately upon jet ejection, as shown in the fluid dynamics video.

%% The format is: \href{URL of video}{name that will appear in the text}

%
\end{document}